\documentclass{iau} 
\usepackage{graphicx}
\usepackage{epstopdf}
\usepackage[T1]{fontenc}
\usepackage[utf8]{inputenc}
\usepackage{lmodern}

\title[Class I methanol masers] 
{Interferometric and single-dish observations of 44, 84 and 95 GHz Class I methanol masers}

\author[Rodr\'iguez-Garza et al.]   
{Carolina B. Rodr\'iguez-Garza$^1$,
Stanley E. Kurtz$^1$,
Arturo I. G\'omez-Ruiz$^2$,
Peter Hofner$^{3,4}$,
Esteban D. Araya$^5$
\and
Sergei V. Kalenskii$^6$
}

\affiliation{$^1$Instituto de Radioastronom\'ia y Astrof\'isica, Universidad Nacional Aut\'onoma de M\'exico, Morelia, Michoac\'an, M\'exico, C.P. 58089\\ email: {\tt ca.rodriguez@irya.unam.mx}  \\[\affilskip]

$^2$Instituto Nacional de Astrof\'isica, \'Optica y Electr\'onica, Luis E. Erro 1, Tonantzintla, Puebla, C.P. 72840, M\'exico \\[\affilskip]

$^3$Physics Department, New Mexico Tech, 801 Leroy Pl., Socorro, NM 87801, USA\\
$^4$Adjunct Astronomer at the National Radio Astronomy Observatory, 1003 Lopezville Road, Socorro, NM 87801, USA\\[\affilskip]

$^5$Physics Department, Western Illinois University, 1 University Circle, Macomb, IL 61455, USA\\[\affilskip]

$^6$Astro Space Center, Lebedev Physical Institute, Profsoyuznaya 84/32, Moscow, 117997, Russia}

\pubyear{2017}
\volume{336}  
\jname{Astrophysical Masers: Unlocking the Mysteries of the Universe}
\editors{A. Tarchi, M.J. Reid \& P. Castangia, eds.}
\begin{document}

\maketitle

\begin{abstract}

We present observations of massive star-forming regions selected from the IRAS Point Source Catalog. The observations were made with the Very Large Array  and the Large Millimeter Telescope to search for Class I methanol masers.
We made interferometric observations of 125 massive star-forming regions in the 44 GHz methanol maser transition; 53 of the 125 fields showed emission. The data allow us to demonstrate associations, at arcsecond precision, of the Class I maser emission with outflows, HII regions and shocks traced by 4.5 $\mu$m emission. 
We made single-dish observations toward 38 of the 53 regions with 44 GHz masers detected to search for the methanol transitions at 84.5, 95.1, 96.7, 107.0, and 108.8 GHz. We find detection rates of 74, 55, 100, 3, and 45\%, respectively. We used a wide-band receiver which revealed many other spectral lines that are common in star-forming regions.


\end{abstract}

\firstsection 
\section{Introduction}

Methanol masers are empirically divided into two classes based on their environments and pumping mechanisms (Batrla et al. 1987; Menten 1991). In particular, the Class I methanol masers arise in molecular gas shocked by outflows in high-mass protostellar objects (HMPOs). These masers were first associated with protostellar outflows by Plambeck \& Menten (1990) and more recently by Cyganowski et al. (2009) by their coincidence with the so-called extended green objects (EGOs), which are regions of shocked gas seen in the 4.5 $\mu$m band of the InfraRed Array Camera (IRAC) on the Spitzer Space Telescope.


Two well-known samples of HMPOs are reported in the literature: the samples of Molinari et al. (1998; hereafter M98) and of Sridharan et al. (2002; hereafter S02). Each sample consists of a collection of 69 HMPOs selected systematically to satisfy specific selection criteria. Half of the sample of M98 contains sources with colors similar to UCHII regions and the other half with colors of deeply embedded sources; 35 of the 69 objects have molecular outflows detected by Zhang et al. (2001; 2005). The S02 sample consists of HMPOs with colors similar to UCHII regions but with 5 GHz flux densities lower than 25 mJy; almost all sources have CO line wings indicative of high velocity gas from molecular outflows (S02, Beuther et al. 2002).

Here, we present interferometric and single-dish observations of both samples to search for Class I methanol maser emision.


\section{VLA surveys of methanol masers \label{vla}}

We made observations of Class I 44 GHz methanol masers with the NRAO\footnote{The National Radio Astronomy Observatory (NRAO) is operated by Associated Universities, Inc., under a cooperative agreement with National Science Foundation.} Karl G. Jansky Very Large Array (VLA) in D configuration toward 69 and 56 HMPOs from the samples of M98 and S02, respectively. The data were observed in Q band (7 mm) using the dual IF mode and fast switching method. One IF of 3.125 MHz bandwith was centered at the 44 GHz methanol transition. The bandwidth was divided in 127 channels providing a spectral resolution of 0.16 km s$^{-1}$ and a velocity coverage of 21 km s$^{-1}$. The observations have an angular resolution of about 2$''$ and a sensitivity of 0.15 Jy. The results of these maser surveys toward the M98 and S02 samples are presented in G\'omez-Ruiz et al. (2016; Survey I) and Rodr\'iguez-Garza et al. (2017; Survey II), respectively.

We find a similar detection rate of 43\% for Class I 44 GHz methanol masers in Surveys I and II. Multiple sources are present in each of the fields from both samples. In general, we note that when an EGO is in the field, the 44 GHz masers seem to favor it, as shown in Figure \ref{fig1}. The spatial coincidence of the 44 GHz masers and the shocked molecular gas supports the idea that these masers may arise from molecular outflows and, presumably, the youngest sources in the field.

Class I methanol masers are thought to be saturated with little time variation over timescales of several years, although this field has not been extensively studied (Leurini et al. 2016). Our observations suggest that variability on these timescales may occur. IRAS 20126+4104 was part of VLA observations of 44 GHz masers made in March 1999 (Kurtz et al. 2004),  March 2007 (Survey I) and August 2008 (Survey II). Four 44 GHz maser components were found clustered toward the NW of the IRAS source and one isolated feature toward the SE of the IRAS central source. Interestingly, the flux density of the isolated feature remains nearly the same during these years while the flux densities of the NW maser group are similar between 1999 and 2007 but are amplified by a common factor of 2 during the observations of 2008. This implies a flux variation on timescales of less than 15 months.

\section{LMT survey of methanol emission}

We observed 38 HMPOs selected from the VLA 44 GHz methanol maser surveys presented in Section \ref{vla}; 19 HMPOs were taken from Survey I and another 19 from Survey II. This sub-sample was observed during the Early Science Phase of the Large Millimeter Telescope (LMT) in 2016. The observations were made with the Redshift Search Receiver (RSR; Erickson et al. 2007), a broad bandwidth spectrometer that covers the frequency range from 73 to 111 GHz, at 31 MHz (100 km s$^{-1}$) spectral resolution. At the time of the observations, the LMT had a usable surface 32-m in diameter, providing a beam size of $\sim 18''$ at 111 GHz. 

The methanol lines within the RSR bandpass are at 84.5, 95.1, 96.7, 107.0, and 108.8 GHz. We find detection rates of 74, 55, 100, 3 and 45\%, respectively.

Many molecular lines are blended in a single LMT channel but in most cases, the methanol emission dominates the channel emission which allows us to identify the spectral line. The low spectral resolution precludes a definitive interpretation of the methanol emission (maser or thermal) but even so, these data serve to identify promising methanol maser candidates for follow-up studies at high spectral resolution.

\begin{figure*}[h!]
\begin{center}
\begin{tabular}{ll}
\includegraphics[width=60mm]{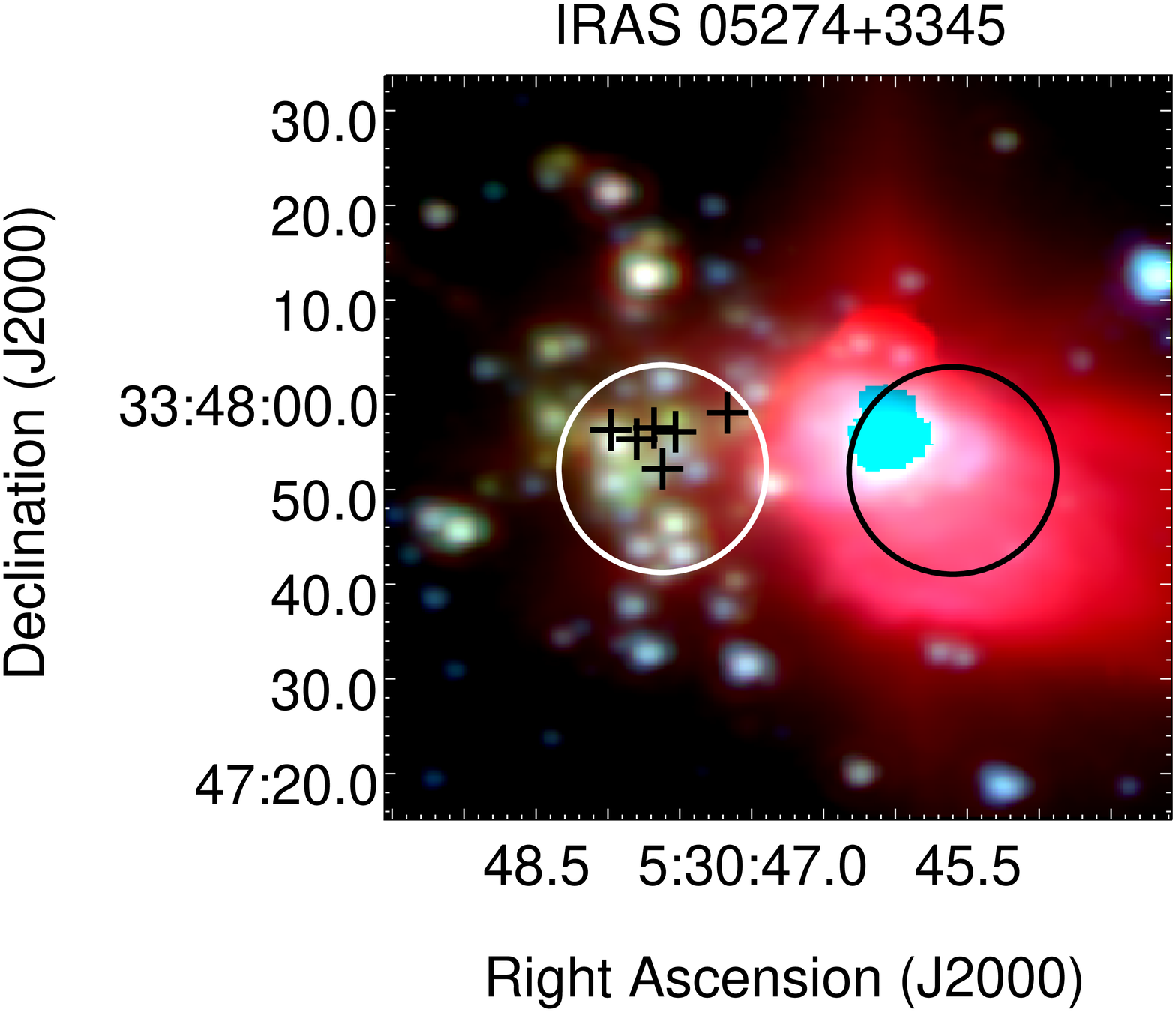} & 
\includegraphics[width=58mm]{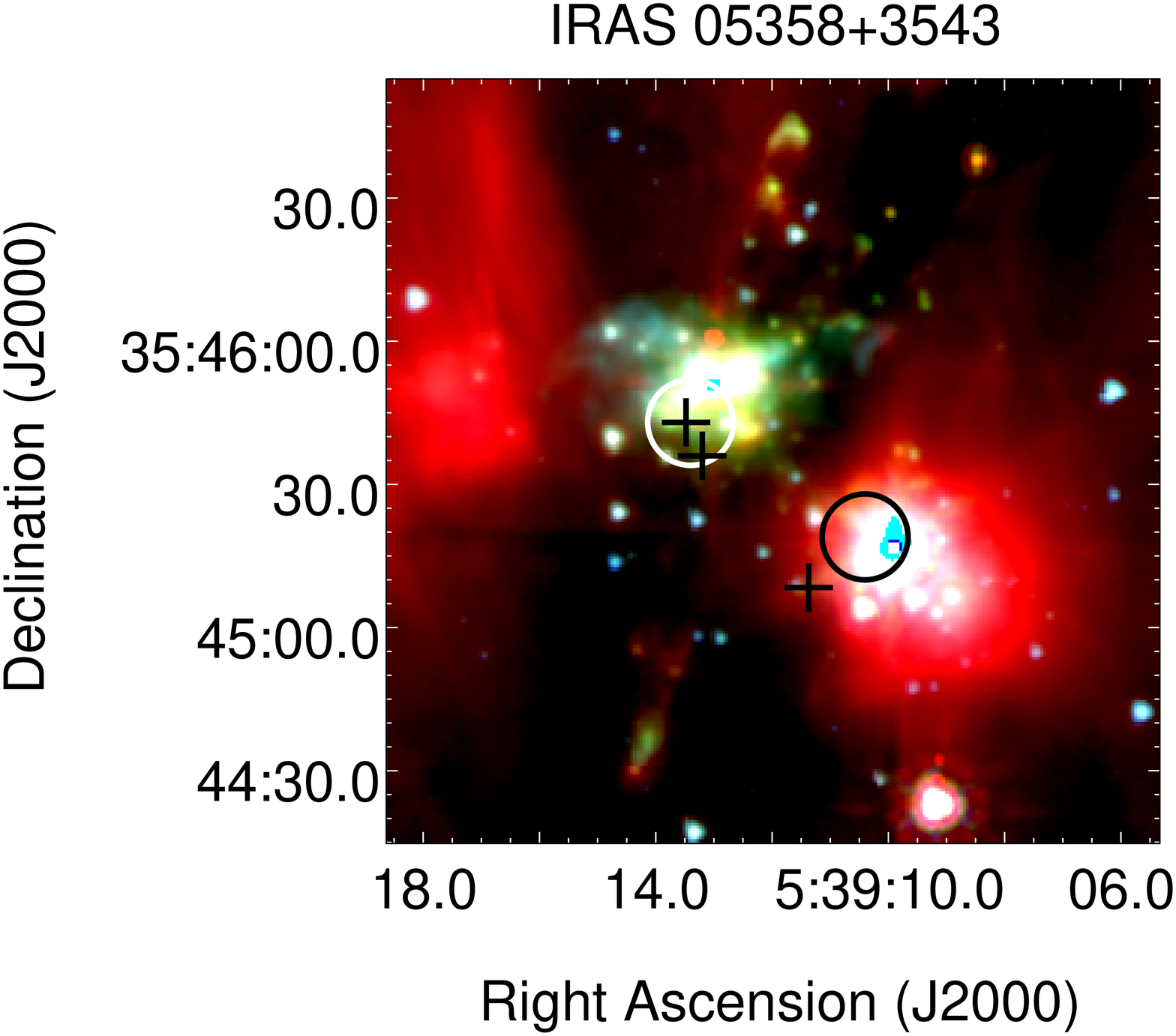} 
\end{tabular}
\end{center}
\caption{Left panel: IRAS 05274+3345 from Survey I. Right panel: IRAS 05358+3543 from Survey II. Spitzer images of the IRAC bands at 3.6, 4.5, and 8 $\mu$m. The crosses are the 44 GHz methanol masers detected in the VLA surveys. The white circles represent the LMT primary beam (18$''$ at 111 GHz) pointed toward the position of the brightest 44 GHz maser (related to shocked gas traced by EGOs) and the black circles represent the LMT primary beam toward the IRAS source.}
\label{fig1}
\end{figure*}

\begin{figure*}[h!]
\begin{center}
\includegraphics[width=14cm, height=5.5cm]{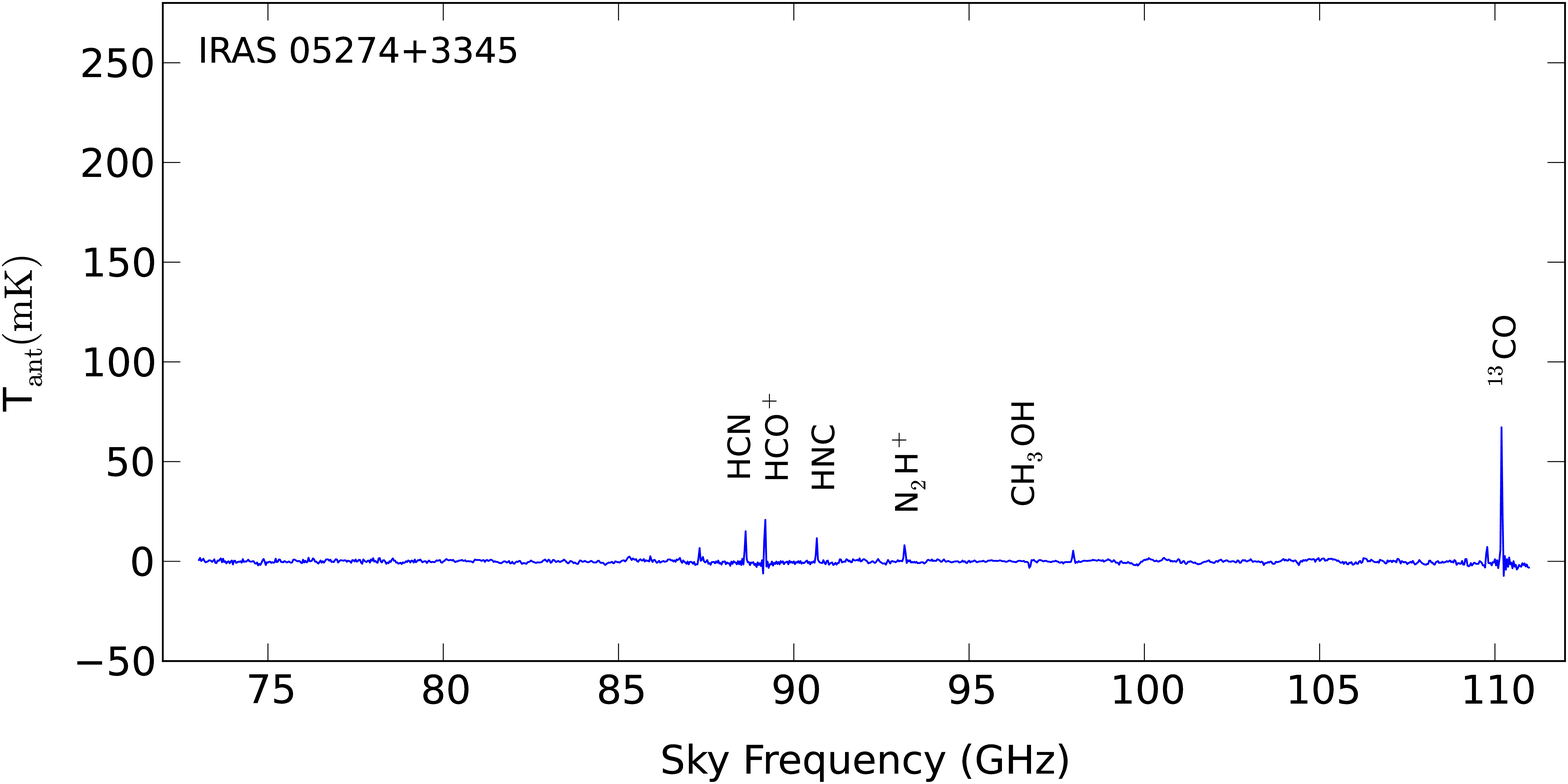} \\ 
\vspace*{0.4 cm}
\includegraphics[width=14cm, height=5.5cm]{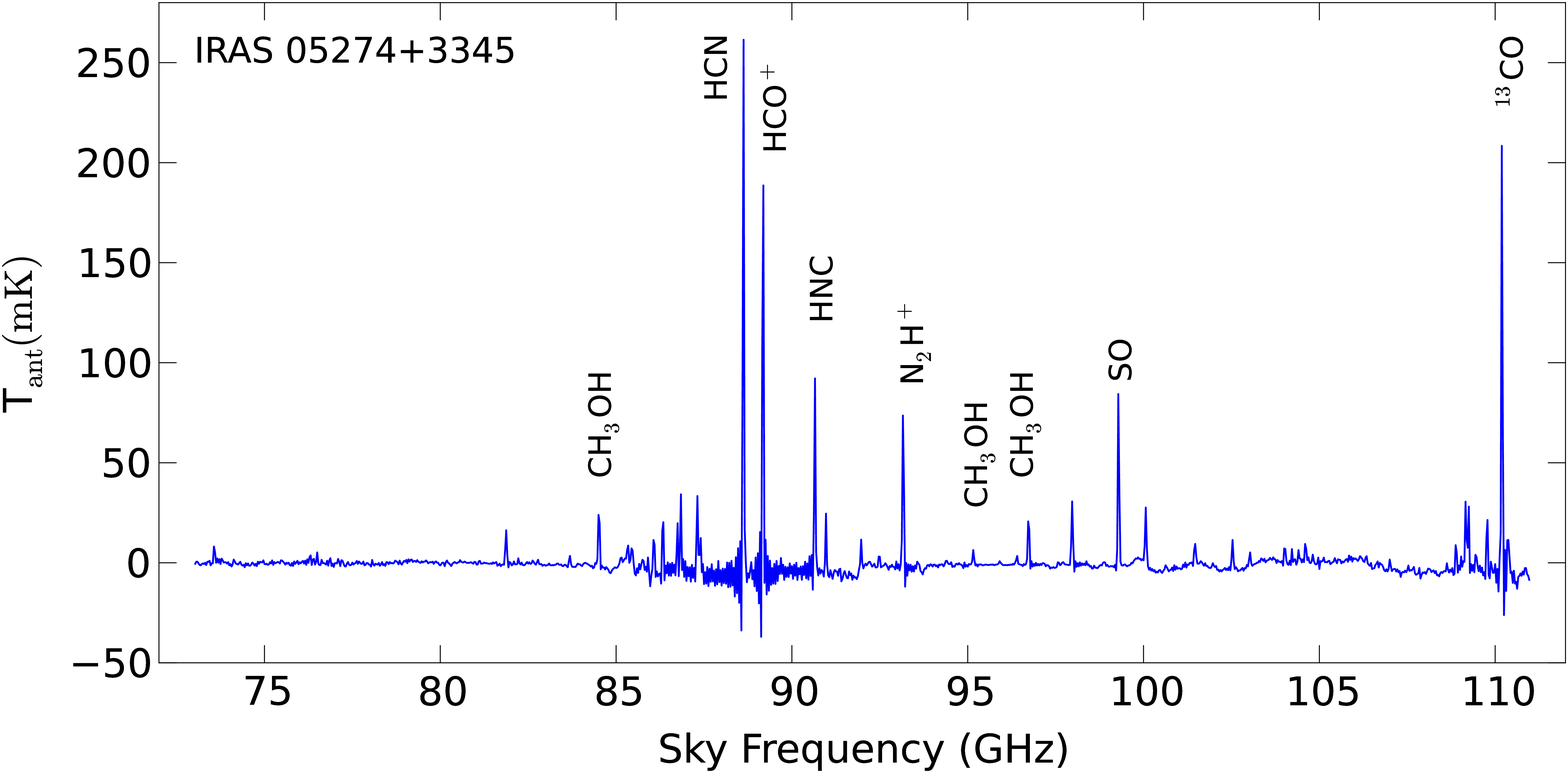}
\caption{Spectra obtained with the LMT in two different pointings toward IRAS 05274+3345. Above: Spectrum taken at the IRAS position. Bottom: Spectrum taken at the position of the brightest 44 GHz methanol maser from Survey I (see Figure \ref{fig1}).}
\label{fig2}
\end{center}
\end{figure*}

The 73-111 GHz band includes many molecular lines commonly found in massive star-forming regions. In order to compare the molecular composition of the shocked regions (EGOs) traced by the 44 GHz masers with the central protostars (IRAS sources), we made observations at both positions. For example, IRAS 05274+3345 and IRAS 05358+3543 show two clusters of star formation in different evolutionary stages (see Figure \ref{fig1}). In both cases, the 44 GHz masers are associated with the youngest cluster, separated by $\sim 30''$ from the IRAS source. We made LMT pointings toward each region to compare their chemical composition. We find that the clusters related to shocked gas are more chemically active than the cluster associated to the IRAS source (see Figure \ref{fig2}).

\section{Conclusions}

Our VLA surveys support the relation of Class I methanol masers with shocked regions around massive protostars. In one case, we found evidence of maser variability, with a time variation of less than 15 months. 

The spectral line survey performed with the LMT in the 3 mm band revealed a rich chemical composition of the shocked gas related to the methanol masers. In the few cases tested, we also found that shocked regions have a greater number of spectral lines than the central regions where the massive protostars are located.

\end{document}